\documentclass[letterpaper]{article}
\usepackage{aaai16}
\usepackage{times}
\usepackage{helvet}
\usepackage{courier}
\usepackage{verbatim}
\usepackage{multirow}
\usepackage{amsmath}
\usepackage{amssymb}
\usepackage{array}
\usepackage{bbold}
\usepackage[dvipsnames]{color}
\usepackage[hyphens]{url}
\usepackage{url}
\usepackage{subcaption}
\usepackage{xspace}
\usepackage{tikz}
\usepackage{afterpage}
\usepackage{booktabs}
\usepackage{mfirstuc}
\usepackage{balance}
\def\year{2016}
\newif{\ifhidecomments}
\hidecommentstrue

\ifhidecomments
    \newcommand{\llee}[1]{}
    \newcommand{\chenhao}[1]{}
    \newcommand{\jack}[1]{}
\else
    \newcommand{\chenhao}[1]{\textcolor{blue}{#1}}
    \newcommand{\llee}[1]{\textcolor{magenta}{#1}}
    \newcommand{\jack}[1]{\textcolor{red}{#1}}
\fi

\newcommand{\addExploreFigure}[2]{\includegraphics[width=#1]{figures/exploration/#2}}
\newcommand{\addTwoLineFigure}[2]{\includegraphics[width=#1]{figures/two_lines/#2}}
\newcommand{\addSpinoffFigure}[2]{\includegraphics[width=#1]{figures/defection/spinoffs/#2}}

\newcommand{\para}[1]{\noindent{\bf #1}}
\newcommand{\figref}[1]{Figure \ref{#1}}

\newcommand{\tableref}[1]{Table \ref{#1}}

\newcommand{\argmin}{\operatornamewithlimits{argmin}}

\newcommand{\defect}{explore\xspace}
\newcommand{\defector}{explorer\xspace}
\newcommand{\adefector}{an \defector}
\newcommand{\Defector}{{\expandafter\makefirstuc\expandafter{\defector}}\xspace}

\newcommand{\defecting}{exploring\xspace}
\newcommand{\defection}{exploration\xspace}

\newcommand{\defsubscript}{e}
\newcommand{\loyalist}{nonexplorer\xspace}
\newcommand{\aloyalist}{a \loyalist}
\newcommand{\Loyalist}{{\expandafter\makefirstuc\expandafter{\loyalist}}\xspace}
\newcommand{\loyal}{\loyalist}

\newcommand{\loyalsubscript}{ne}

\newcommand{\modification}{affix\xspace}
\newcommand{\amodification}{an affix\xspace}
\newcommand{\modifications}{affixes\xspace}
\newcommand{\modified}{{\modification}ed\xspace}
\newcommand{\Modification}{{\expandafter\makefirstuc\expandafter{\modification}}\xspace}
\newcommand{\unmodified}{{un{\modified}}\xspace}

\newcommand{\hrc}{highly related communities\xspace}
\newcommand{\HRC}{Highly Related Communities\xspace}
\newcommand{\hrcsingular}{highly related community\xspace}

\newcommand{\spinoff}{spinoff\xspace} %

\newcommand{\Spinoff}{{\expandafter\makefirstuc\expandafter{\spinoff}}\xspace}

\newcommand{\submission}{post\xspace}
\newcommand{\Submission}{{\expandafter\makefirstuc\expandafter{\submission}}\xspace}

\newcommand{\founder}{early participant\xspace}

\newcommand{\communityname}[1]{{\small\sf #1}\xspace}
\newcommand{\categoryname}[1]{{``#1''}\xspace}
\newcommand{\communitypair}[2]{\communityname{#1} vs. \communityname{#2}}
\newcommand{\modifier}[1]{{\em #1}\xspace}
\definecolor{defectorcolor}{RGB}{255,121,108}
\definecolor{loyalcolor}{RGB}{106,121,247}
\newcommand{\metricname}{exploration effect\xspace}
\newcommand{\Metricname}{Exploration effect\xspace}
\newcommand{\ametricname}{an exploration effect\xspace}

\newcommand{\namecite}[1]{\citeauthor{#1} \shortcite{#1}\xspace}

\title{
Science, AskScience, and BadScience:\\
On the
Coexistence
of Highly Related Communities
}

\author{
Jack Hessel \and Chenhao Tan \and Lillian Lee\\
Department of Computer Science\\
Cornell University\\
\texttt{\{jhessel, chenhao, llee\}@cs.cornell.edu}
}

\begin{document}
\maketitle
\begin{abstract}
When large social-media
platforms allow users to easily form and self-organize into interest
groups, highly related communities can arise. For example, the Reddit
site hosts not just a group called
\communityname{food}, but also \communityname{HealthyFood},
\communityname{foodhacks}, \communityname{foodporn}, and \communityname{cooking}, among others.
Are these \hrc created for similar classes of
reasons
(e.g., to focus on a subtopic, to create a place for allegedly more ``high-minded'' discourse, etc.)?
How do users allocate
attention between such close alternatives when they are available or
emerge over time?  Are there different types of relations between
close alternatives such as sharing many users vs. a new community
drawing away members of an older one vs. a splinter group failing to
cohere into a
viable separate community?  We investigate the interactions between
highly related communities using
data from \texttt{reddit.com} consisting of
975M posts and comments spanning an 8-year period.  We identify a set
of typical \modifications that users adopt to create \hrc and build a
taxonomy of \modifications.
One interesting
finding
regarding users' behavior is: after a newer community is created, for
several types of highly-related community pairs, users that engage in
a newer community tend to be \emph{more active} in their
\emph{original}
community than users that do
not explore, even when controlling for previous level of engagement.
\end{abstract}

\section{Introduction}

Social networks are in constant flux, with new communities forming and
old communities dying over time.  On websites such as Facebook and
Reddit, users have complete freedom to create communities at their own
discretion.  This has led to a very large number of communities
arising organically from user initiative,
for a variety of reasons.
One reason is to create
divisions that
satisfy
the need to better organize discussions;
in fact, community design theory argues that ``a growing Web community needs subdivisions which might be represented as towns, neighborhoods, topics, categories, conferences, or channels, depending on your metaphor'' \cite{Kim:2000:CBW:518514,Jones:ElectronicMarkets:2010}.
Or, new groups can develop because of religious, political, or other
schisms;
online examples include groups whose very names attempt to connote
superiority to others, e.g., the
subreddits \communitypair{trueatheism}{atheism}.
Other reasons surely exist.
The tremendous reach of modern social
media
provides researchers much greater
data to examine these social processes at scale.

\begin{table}[t]
\centering
\caption{The 10 most common  Reddit group-name \modifications.}
\label{tb:modifications}
\begin{tabular}{llr} \toprule
\Modification & Example & \# Pairs \\ \midrule
\modifier{s} & \communityname{auto}, \communityname{autos} & 63 \\
\modifier{porn} & \communityname{space}, \communityname{spaceporn} & 26  \\
\modifier{circlejerk} & \communityname{hiphop}, \communityname{hiphopcirclejerk} & 23 \\
\modifier{ask} & \communityname{science}, \communityname{askscience} & 21 \\
\modifier{shitty} & \communityname{ideas}, \communityname{shittyideas} & 17 \\
\modifier{music} & \communityname{running}, \communityname{runningmusic} & 17 \\
\modifier{help} & \communityname{tech}, \communityname{techhelp} & 11\\
\modifier{2} & \communityname{dota}, \communityname{dota2} & 9\\
\modifier{true} & \communityname{atheism}, \communityname{trueatheism} & 9 \\
\modifier{learn} & \communityname{math}, \communityname{learnmath} & 9\\

\bottomrule
\end{tabular}
\end{table}

An interesting
and frequently occurring version of the
group creation process
 is that a new concept or
culture may gain in popularity and,
in a meme-like fashion,
 draw users
to create a new community by using that concept as \emph
{\amodification}\footnote{
An affix is either a prefix or a suffix.} of
their community name.
For example, on Facebook, after the creation of the \communityname{OMG
Confessions} group,
anonymous confession pages with names combining a college with the
word \modifier{confession} or \modifier{confessional}
proliferated to the degree
that one can
now find a confession
page for almost every university campus.
(\citeauthor{Birnholtz:2015:WSV:2702123.2702410} \shortcite
{Birnholtz:2015:WSV:2702123.2702410}
examine what kind of questions people
ask
on such pages.)
\tableref{tb:modifications} shows some examples from Reddit: the second
column shows pairs of subcommunities where the name of one is a modified
form of the other (ignore the third column for now).\footnote{
An additional, whimsical example
from Reddit is \modifier{random\_acts\_of\_}, indicating people asking for or sending free things
to others.
Instantiations include \communityname{random\_acts\_of\_pizza}, \communityname{random\_acts\_of\_amazon}, and \communityname{random\_acts\_of\_books}.
\citeauthor{althoff+al:14a} \shortcite{althoff+al:14a} used
\communityname
{random\_acts\_of\_pizza} to study
effective ways to ask for a
favor.
}
In this work, we investigate  \hrc that are based on \modifications.
An understanding of these \hrc may help community organizers
identify subtopics in a community
and create an appropriate subdivision to cultivate focused discussions,
or monitor subgroups that potentially feel marginalized or underserved, and
decide whether to change community norms or create a dedicated community for that subgroup.

Despite the ubiquity
of such \modifications,
and their appeal as easily-identifiable
(albeit sometimes imperfect)
instances of the important phenomenon of \hrc,
little is known about canonical \modifications and the activity in the resultant \hrc.
For instance, are neighborhoods, topics, and channels enough to capture all
possible \modifications? Are there classes of \modifications that are
generally applicable?
Perhaps different \modifications behave in different ways.
Moreover, once a \hrcsingular is created, how does it interact with the existing
community? Will it overtake it? Will the two share the same
user base?
One of our goals is to analyze user behavior in the existing community
\emph{after they participate in the new community.}

\para{Organization and contributions.}
In this paper, we construct a dataset from Reddit and present the first large-scale study on the coexistence of \hrc.
Details about the dataset are introduced in ``Dataset Description''.

Our first contribution is to characterize the space of {\em \modifications}.
We build a taxonomy of common \modifications that users adopt to create \hrc.
For instance, we identify a category of \categoryname{parody}
\modifications (\modifier{circlejerk}, \modifier{shitty},
\modifier{funny}, \modifier{lol}, \modifier{bad}). This category
generally shares the same
user base with its corresponding \unmodified
community. On the other hand, we identify a category of
\categoryname{derivative} \modifications (\modifier{meta}, \modifier{anti},
\modifier{srs}, \modifier{post}, \modifier{ex}) that likely attract
different user bases.  Surprisingly, a non-trivial fraction of
\modified communities exist before the \unmodified ones.  Also, an
interesting class of {\em \spinoff} communities arises where
{\founder}s in the new community
come from the existing community.

Our second contribution is to introduce a framework for analyzing
users who try out \spinoff communities (dubbed ``{\defector}s'') and
comparing them to ``{\loyalist}s'' who never leave the original
subreddit.
We make the surprising observation that in multiple classes of
\modifications, users who explore \spinoff communities are \emph{more
  active in the original communities after \defecting}
when compared to similarly active users who never tried the
alternative.  This resonates with the findings in
\citeauthor{tan2015all} \shortcite{tan2015all} that users who
``wander'' to different (potentially completely unrelated) groups tend
to stay active longer on the site as a whole. Our observations
may
suggest that \spinoff communities generally serve a complementary
rather than competitive role in multi-community settings.
Finally, we summarize related work and offer some concluding
thoughts.

\section{Dataset Description}

Our starting point for understanding \hrc, \modifications,
{\spinoff}s, {\loyalist}s, and {\defector}s is an examination of
\emph{topically related communities}.
As such,
we
compile
a dataset from \texttt{reddit.com}, a site
where users are
allowed
to create communities called subreddits at
their discretion. Users can name the subreddits that they create so
that
like-minded people can identify
them effectively.
As a result of unmoderated
creation and limitless naming possibilities, there are a wide variety
of subreddits on Reddit, e.g., \communityname{funny}, \communityname{worldnews}, \communityname{politics}, \communityname{IAmA},
\communityname{todayilearned}, etc. On these subreddits, users submit
link-based {\submission}s or text-based {\submission}s, comment on
others' {\submission}s, and up/down vote {\submission}s and comments.
We construct a dataset that includes all
 activities on Reddit from its inception until 2014,
an 8-year period, by combining two data sources:
a \submission dataset that was
organized
in
\namecite{tan2015all}, and
all comments data
extracted by Jason Baumgartner.\footnote{Information is available at \url{https://pushshift.io}.
The dataset in \namecite{tan2015all} was also originally extracted by Jason Baumgartner.
}
We focus on communities that are active
and that enjoy a reasonable number of
users.
Specifically, we require all communities to include at least 300
unique users that made {\submission}s. This left
us with just under 5.7K
communities.
\tableref{table:dataset_stats} presents basic statistics of this
dataset.\footnote{The statistics reported here include posts and
  comments made by users who deleted their accounts and banned
  accounts.} The metadata for
the
   Reddit conversation trees that we used
here is available for
download.%
\footnote{\url{http://goo.gl/sHUfhC}}

\begin{table}[t]
\centering
\caption{Summary statistics for our Reddit corpus.
  {\Submission}s are from
  \citeauthor{tan2015all} \shortcite{tan2015all} and include all {\submission}s on Reddit from its inception in 2006 to February, 2014.
All comments on these posts up until November 2014 were drawn from Jason Baumgartner's comment dataset.
  }
\begin{tabular}{p{2in}r} \toprule
Data type & count \\
\midrule
Subreddits & 5,692 \\
{\Submission}s & 88M\\
Comments & 887.5M\\
\bottomrule
\end{tabular}
\label{table:dataset_stats}
\end{table}

As discussed in the introduction,
user-defined subreddit names are an important indicator of
relationships between \hrc (e.g., \communitypair{food}{HealthyFood}).
We first retrieve
all possible pairs of communities where one community name is the
other's suffix or prefix, ignoring case (\communityname{food} is the suffix of
\communityname{HealthyFood}, ignoring case). We refer to the difference between the
names in a pair as the {\em \modification}. For instance, \modifier{healthy} is the
\modification in the pair \communityname{food} vs. \communityname{HealthyFood}.  There are around 4K
such pairs over our dataset.

Using common affixes as a starting point allows us to discuss the
the space of possible \hrc. For example, this framing allows
us to make statistical observations about all
pairs with
\modifier{healthy} or
\modifier{true}
as \modifications.
Note that %
we
omit some
interesting \hrc pairs by focusing on \modified pairs.
One example is
\communityname{TwoXChromosomes}, a very popular ``subreddit ... intended
for women's perspectives,'' and \communityname{TrollXChromosomes},
its satirical counterpart.

\para{Identifying topically related communities.}  Unsurprisingly, not
all pairs of communities identified through \modifications are
actually \hrc. An example is ``ru'' and ``rum;'' the first one is a
Russian community while the second one is about the liquor.
In order to quantify subreddit similarity, we compute the content
similarity between pairs of communities. As suggested in
\namecite{Singer:Www14:2014},
subreddits can focus on text {\submission}s in addition to link-based
{\submission}s.
Therefore, we employ a method that can account
for either link-dominant or text-dominant subreddits. Specifically, we
use Jaccard similarity between the set of links to capture similarity
based on links,\footnote{Jaccard similarity is defined as $\frac{A
    \cap B}{A \cup B}$, where $A$ and $B$ are the set of links from
  two subreddits respectively.} and use Jensen-Shannon divergence
between topic distributions (derived from a topic model trained on
6.6M text {\submission}s) to capture similarity based on
text,
following \namecite{hessel2015vegans}.
Since these two metrics are not comparable by raw value,
we compute the full background distribution based on all
1.62M possible pairs of the 5.7K communities in our dataset
and compute the percentile of each \modification pair in each
distribution.

We consider a pair of communities to be topically related if either
link similarity is above
the 90th percentile \emph{or} topical similarity based on text
is above the 90th percentile.  Accounting for our definition of topical
similarity yields just over 1.7K pairs from our original set of 4K.

The last step of our preprocessing is to identify generalizable
\modifications that are commonly used in these \hrc.  We count the
frequency of \modifications and keep \modifications that occur at
least three times, so that all \modifications in the final dataset
carry a general
meaning (it is not possible to make general statements about
\modifications that only occur once).  This step brings us to 99
\modifications and 572 pairs of highly related communities
distributed between them.

\begin{table}[t]
\centering
\caption{A taxonomy of \modifications.}
\label{tb:taxonomy}
\begin{tabular}{p{0.8in}p{2.2in}} 
\toprule
\multicolumn{2}{c}{Adjective-like}\\
\midrule
``better'' & true, plus\\
``parody'' & circlejerk, shitty, funny, lol, bad\\
``derivative'' & post, ex, meta, anti, srs \\
``genre'' & classic, fantasy, indie, folk, casual, dirty, classic, metal, academic, 90s, free, social\\
``nsfw'' & nsfw\_, nsfw, asian, trees, gonewild, gw, r4r, tree\\
\midrule
\multicolumn{2}{c}{Verb-like}\\
\midrule
``learning, improvement'' & ask, help, learn, advice, hacks, stop\\
``action'' & exchange, randomactsof, trade, trades, classifieds, market, swap, random\_acts\_of\_, requests, invites, builds, making, mining, craft\\
\midrule
\multicolumn{2}{c}{Noun-like}\\
\midrule
``place'' & uk, reddit, chicago, us, dc, steam, canada, american, boston, android, online, web\\
``medium'' & porn, pics, music, memes, videos, vids, comics, apps, games, gaming, game\\
``subject'' & science, news, dev, servers, tech,  tv, guns, recipes, city, u, college, man, girls\\
\midrule
\multicolumn{2}{c}{Minor}\\
\midrule
``equivalent, competition'' & s, al, ing, the, alternative\\
``generation'' & 2, 3, 4, 5 \\
``modifier'' & ism, n, an\\
\bottomrule
\end{tabular}
\end{table}

\section{Characterizing \modifications}

The goal of this section is to explore the types of canonical
\modifications users on Reddit utilize. To accomplish this
exploration, we first build a taxonomy of common \modifications to
better understand their basic properties and relationships.
Next, we explore the temporal characteristics of the
pairs.
In general, we observe
an accelerating culture of creating \hrc, meaning that \hrc are being
created at increasing rates.
We also observe that, in most cases, the \modified community in a pair
was created after the \unmodified one, even though there is a
non-trivial fraction that went the other way, e.g.,
\communityname{ukpolitics} and \communityname{uspolitics} both existed
before \communityname{politics}.
We further explore whether the newer community ``overtakes'' the older one
in
popularity.  We then offer
potential rationales
that may help explain
the surprising finding
that a \emph{quarter of the newer communities are more active.}  The
final characteristic that we examine is whether the newer community
actually shares a user base with the older one, at least when the new
one is forming.
Despite the high similarity both in
community name and in content, almost half of newer subreddits in
pairs are not, in fact, born out of their older partners.
\subsection{The space of \modifications}
In order to achieve a basic understanding of what canonical \modifications
users adopt to create new communities,
we first build a taxonomy of the 99
\modifications from the dataset section
in \tableref{tb:taxonomy}.

We start with a coarse structure based on part-of-speech.
Among the adjective-like, the largest category is based on \categoryname{genre}, e.g., \communitypair{rock}{classicrock}.
Some other very interesting classes also arise:
\categoryname{better}, which indicates a certain level of superiority (e.g., \communitypair{atheism}{trueatheism});
communities dedicated to \categoryname{parody} where users are likely aware of the culture in the \unmodified one (e.g., \communitypair{history}{badhistory});
and \categoryname{derivative}, which  probably attracts
a
 very different audience (e.g., \communitypair{war}{antiwar}).
In fact, \modifier{anti} and \modifier{meta} can be recursive, e.g.,
\communityname{jokes}, \communityname{antijokes} and \communityname{antiantijokes}.

Among the verb-like \modifications, a class of self-improvement or learning
communities exists, e.g.,
\communitypair{programming}{learnprogramming}.
In \categoryname{actions}, there are many exchange related \modifications, including \modifier{trades} (e.g., \communitypair{pokemon}{pokemontrades}) and \modifier{swap} (e.g., \communitypair{scotch}{scotchswap}).
Altruistic behavior signified by \modifier{random\_acts\_of\_} (e.g., \communitypair{pizza}{random\_acts\_of\_pizza}) has been studied specifically in
\namecite{althoff+al:14a}.

The noun-like \modifications closely match the conceived metaphor of splitting space in community design theory \cite{Kim:2000:CBW:518514}.
Indeed, we see a group of \modifications based on \categoryname{place}, such as \modifier{uk} (e.g., \communitypair{politics}{ukpolitics}).
\categoryname{Medium},
named for the
\emph{medium} of the content, e.g., \modifier{pics}, \modifier{vids}, etc.
 is another common
 category, including \modifier{videos} (e.g., \communitypair{cat}{catvideos}).
The last one is based on \categoryname{subject}, such as \modifier{recipes} (e.g., \communitypair{vegan}{veganrecipes}).
Noun-like \modifications are probably used to encourage better discussions. These communities do not necessarily share similar users;
 e.g., people who are interested in \communityname{veganrecipes} may not be vegans; people who are invested in \communityname{ukpolitics} may not care about politics in general.

Surprisingly, there is a class of relatively minor changes that can
cause community pairs to differ significantly.  An example of
\categoryname{modifier} is \modifier{ism}, as in
\communitypair{vegetarian}{vegetarianism}, which
align in topic but likely attract
different
people.
Another interesting class is \categoryname{equivalent}.
One example is \communitypair{wallpaper}{wallpapers}:
these two subreddits have indistinguishable
(to the authors)
content and thousands of
members each, yet the moderator sets are disjoint, and
neither mentions the other in their respective extensive
and overlapping
lists of related
subreddits.
In cases like this, the newer community may be created
without knowing about the older one,
although in other cases there may be known prior interactions; for example,
\communityname{Politic} was created because some users do not
like the rules in \communityname{politics}.

Although some decisions in
our
taxonomy are arbitrary, we consider it
useful and meaningful to get an overall sense of possible
\modifications. All \modifications we consider seem to be
generalizable changes that one can make with some community name to
obtain another community name.

\begin{figure}[t]
    \begin{subfigure}[t]{0.23\textwidth}
        \addExploreFigure{\textwidth}{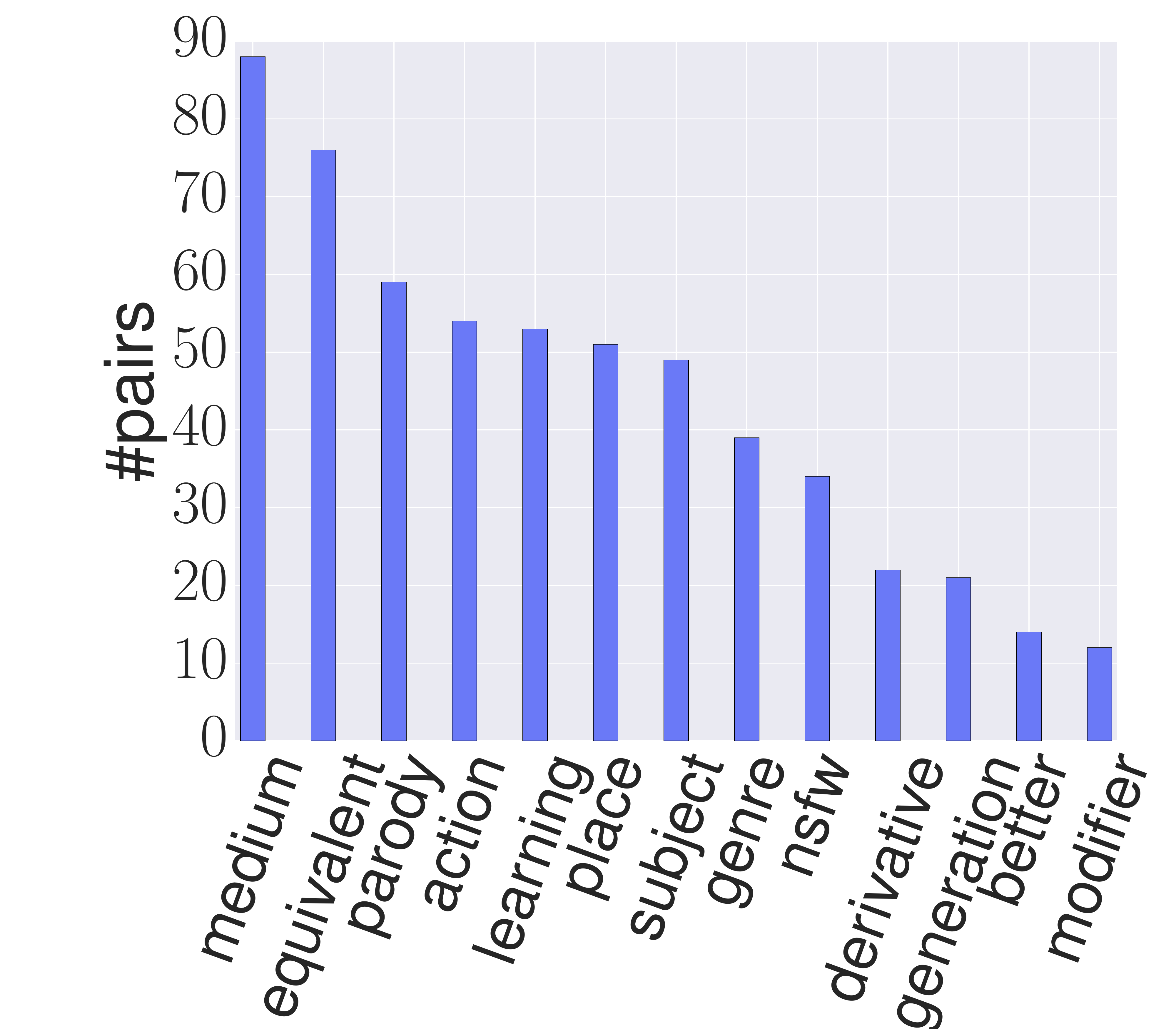}
        \caption{\#Pairs by category.}
        \label{fig:modification_frequency}
    \end{subfigure}
    \hfill
    \begin{subfigure}[t]{0.23\textwidth}
        \addExploreFigure{\textwidth}{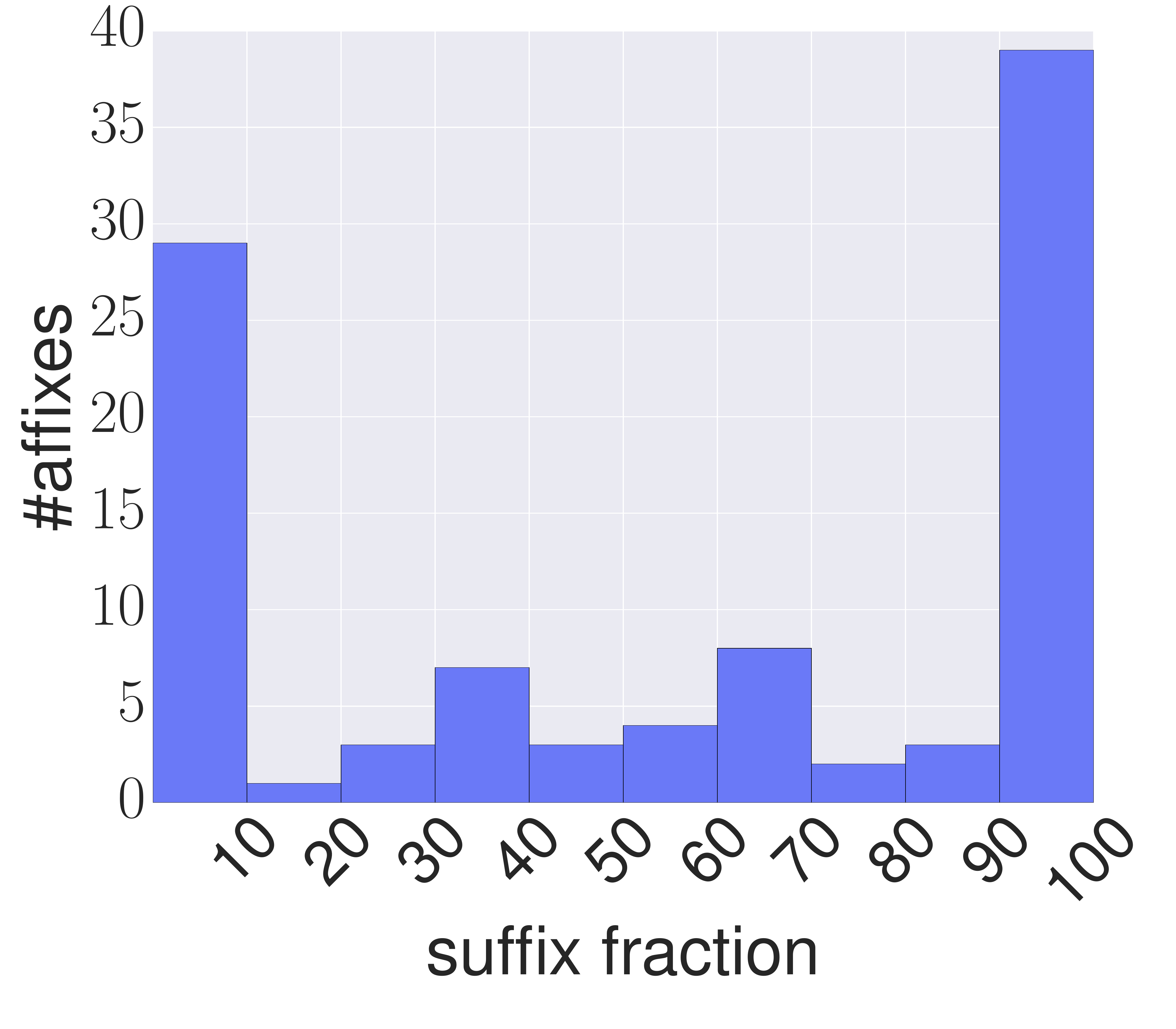}
        \caption{Histogram of fraction as suffixes for all \modifications.}
        \label{fig:suffix_fraction_dist}
    \end{subfigure}
    \caption{
        (a)
        Medium is the most frequent \modification, while modifier is the least.
        (b) Two distinct types of \modifications exist: suffix-dominant and prefix-dominant.
    \label{fig:modification_space}}
\end{figure}

\para{Frequency of \modifications.}
Next, we examine the frequency of \modifications.
\tableref{tb:modifications} presents the 10 most common
\modifications and \figref{fig:modification_frequency} shows
the frequency by category.
The most common \modification is simply the character
\modifier{s}, which suggests that it is perhaps common for
``redundant''
communities to be created.
The most common category is \categoryname{medium}
with 88 pairs,
while the least one is \categoryname{modifier} with 12 pairs.  There
is
some variation in frequency within each category.  For instance, one
interesting observation is that although \modifier{porn} and
\modifier{pics} both fall in \categoryname{medium} and indicate a
related
picture-driven community, \modifier{porn}
has more than 4
times more \hrc (33 vs 7).

\para{Position of \modifications
in community name.}
As shown in \figref{fig:suffix_fraction_dist},
most
but not all
affixes are either
suffix-dominant or prefix-dominant.
Overall, \categoryname{generation}, \categoryname{medium}, and \categoryname{modifier} tend to be used as suffixes, while
\categoryname{genre}, \categoryname{derivative}, and \categoryname{place} are usually used as prefixes.  \categoryname{parody} and \categoryname{nsfw} can be used either way, for example, \modifier{funny} in
\communitypair{videos}{funnyvideos} and \communitypair{Guildwars2}{guildwars2funny}.

\subsection{Temporal Relationships within Pairs}

\begin{figure}[t]
    \begin{subfigure}[t]{0.23\textwidth}
        \addExploreFigure{\textwidth}{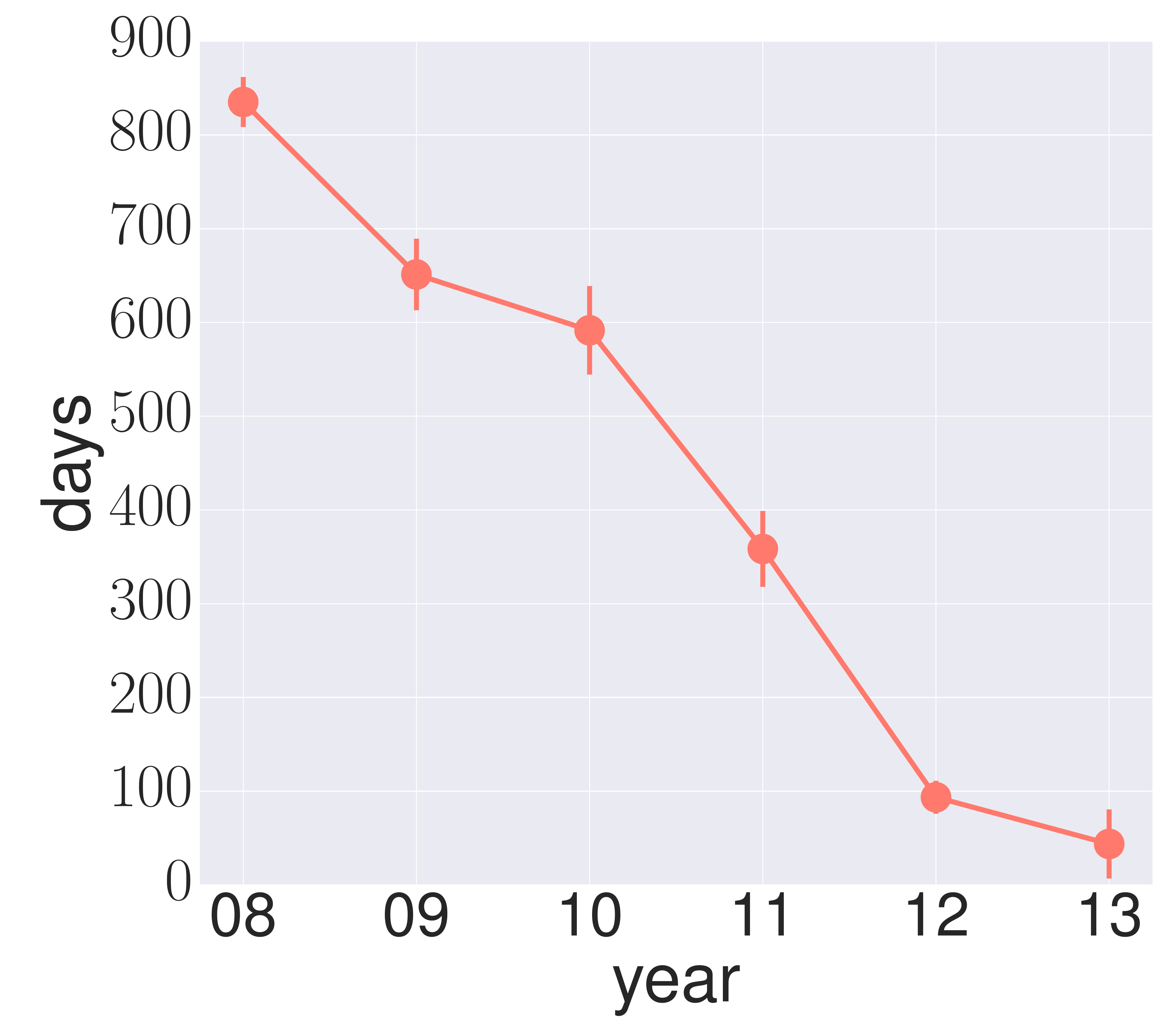}
        \caption{Average gap for pairs grouped by the creation year of the older community.}
        \label{fig:gap_year}
    \end{subfigure}
    \hfill
    \begin{subfigure}[t]{0.23\textwidth}
        \addExploreFigure{\textwidth}{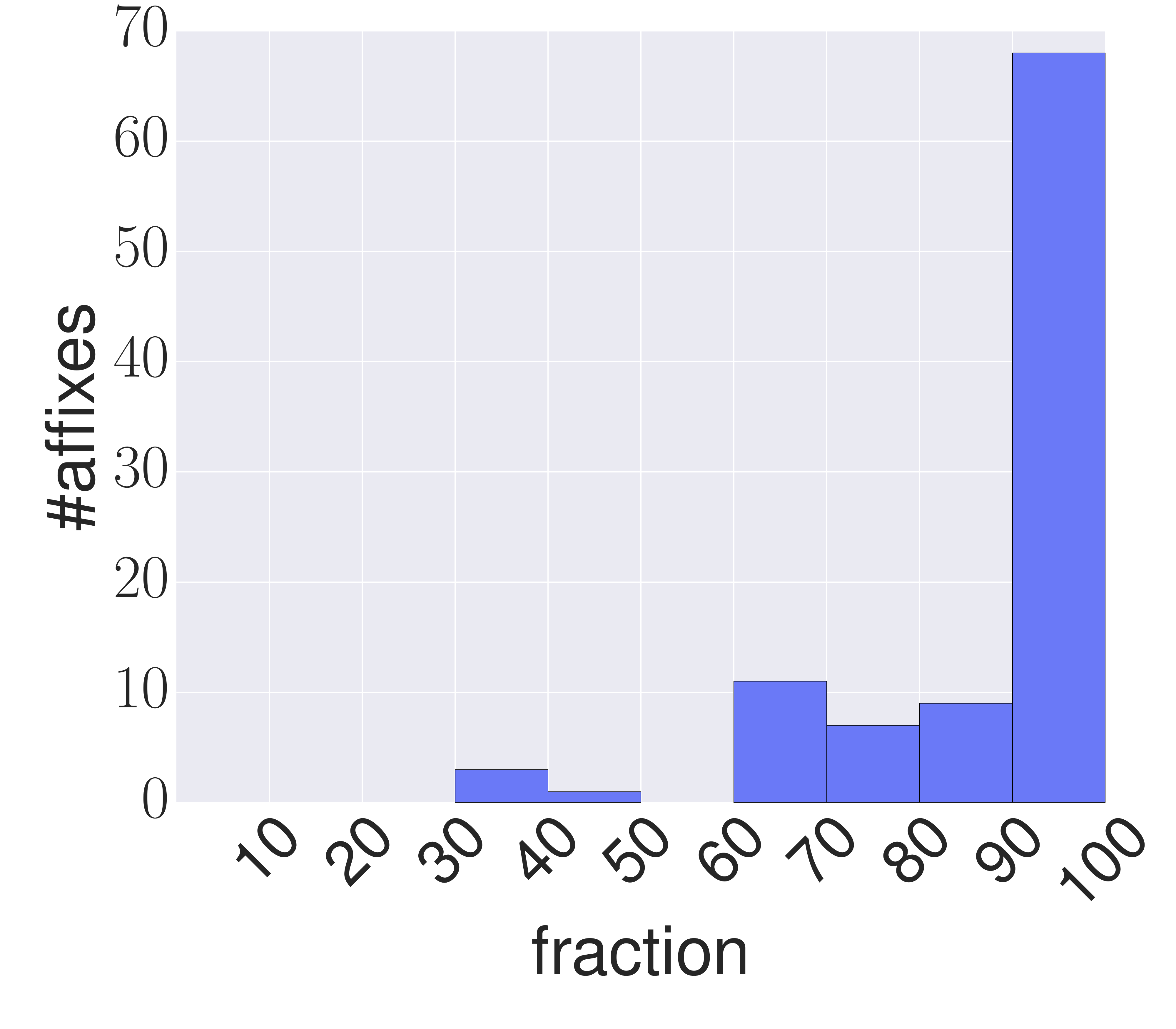}
        \caption{Histogram of
        fraction where the \modified community was created later.}
        \label{fig:order_dist}
    \end{subfigure}
    \caption{
   (a) The newer related community is created with shorter and shorter gap. (b)
   For most \modifications, the \modified community was created later, though there
   are many counterexamples.
    \label{fig:temporal}}
\end{figure}
It is always possible to determine which community in a pair was created earlier.
The first characteristic that we examine is the gap between the creation time of two communities in a pair.
The overall average gap is 749 days since 2008,
when users on Reddit were first allowed to create their own
communities.
If we compute the average gap grouped by the creation
year of the older community, in \figref{fig:gap_year} we see a
consistent trend that the newer community is created with a shorter
and shorter gap over time.
This suggests that there may be
an accelerating culture of
creating \hrc
over time,
or that as there are more users
on Reddit, \modified communities
arise more quickly.

\para{For most \modifications, the community with the \modification
  was newer.} We further examine whether the newer community within a
pair is the \modified one.
This is indeed the case in 86\% of our pairs.
However, if we change our focus from pairs to \modifications, we
find that for 33\% of the \modifications, there was at least
one instance where the
\modified version was actually \emph{created before} the original (see
\figref{fig:order_dist}).

The four \modifications for which the \modified version of the
community more often exists first
are \modifier{ing}, \modifier{al}, \modifier{ism} and \modifier{s}; these are mostly
in the \categoryname{equivalent/competition} class in
\tableref{tb:taxonomy}.
As a result, we observe phenomena like different communities
focusing on exactly the same thing (e.g.,
\communitypair{wallpaper}{wallpapers}) or two communities eventually
deciding to explicitly merge into one (e.g.,
\communitypair{wedding}{weddings}).
Communities with different foci
but similar names might also fall into this category,
such as \communitypair{vegetarian}{vegetarianism}.

These four \modifications do not cover all possible
cases where the \modified was created earlier.  For instance,
\communityname{twincitiessocial} was created before
\communityname{twincities}.
\subsection{Does the New Overtake the Old?}

Another important characteristic is how active the newer community is
compared to the older one after its inception.

\begin{figure}[t]
    \begin{subfigure}[t]{0.23\textwidth}
        \addExploreFigure{\textwidth}{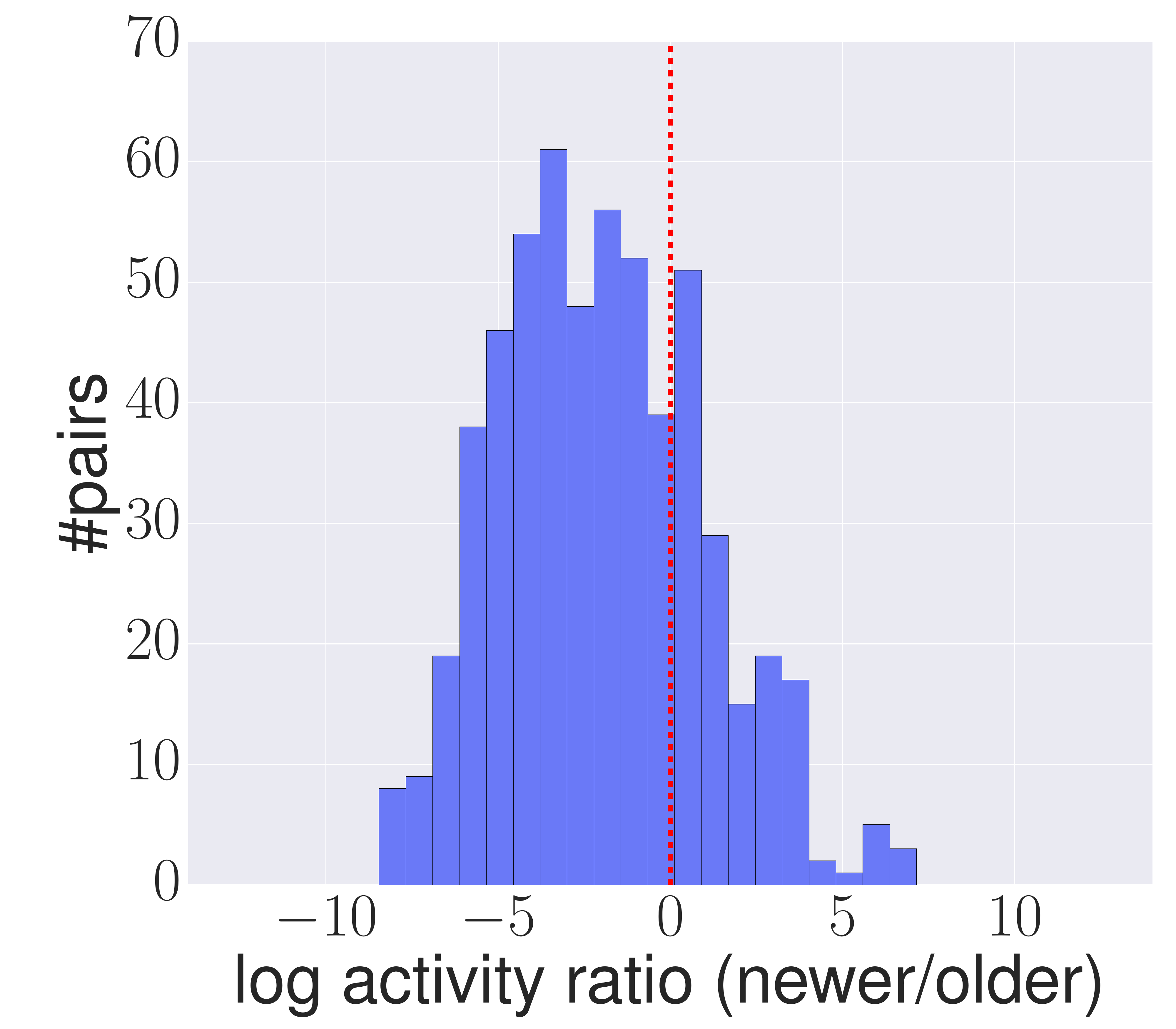}
        \caption{Histogram of log activity level ratio between the newer community and the old one.}
        \label{fig:activity_dist}
    \end{subfigure}
    \hfill
    \begin{subfigure}[t]{0.23\textwidth}
        \addExploreFigure{\textwidth}{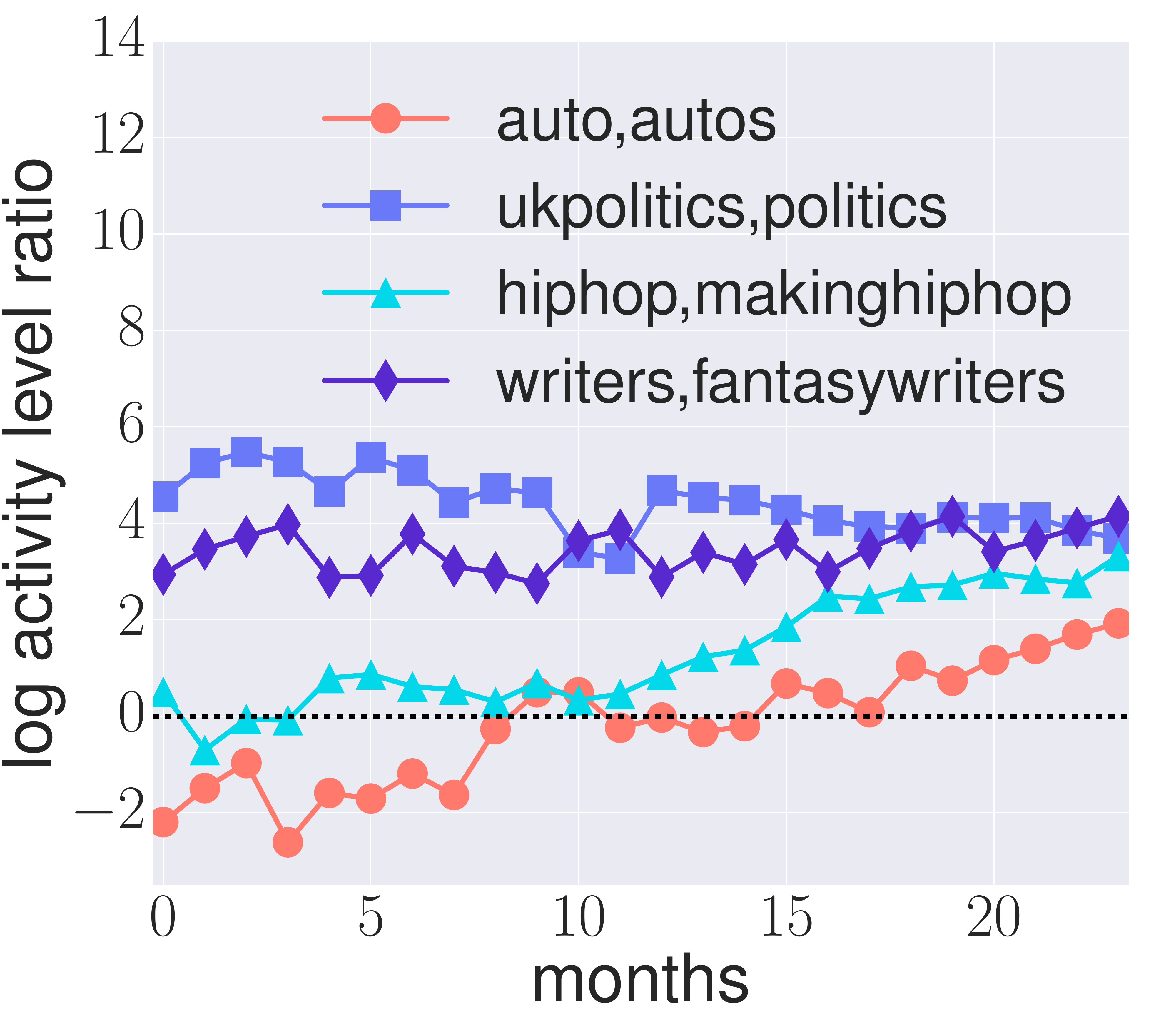}
        \caption{Case study on pairs in which the newer one has more
          activity, where activity is binned on a month-to-month
          basis.}
        \label{fig:activity_case}
    \end{subfigure}
    \caption{
   (a) The older community tend to have a higher level of activity. (b) Examples of different reasons that the newer one can have more activity.
   It shows how the log activity level ratio changes over time since the newer one was created in the first two years.
    \label{fig:activity_level}}
\end{figure}

\para{Newer communities tend to be less active, but in a quarter of
  pairs, the newer one is more active.}  We compute the log ratio in
activity level (the total number of comments plus the total number of
{\submission}s)
between the newer community
and the older community with add-one smoothing,
only considering actions \emph{after the newer community was created}
so that we compare pairs during the same time period. According to
this metric, a positive value means more activity in the newer
community and a negative value means less activity in the newer
community.
\figref{fig:activity_dist} demonstrates there is a trend that \modified
versions of communities tend to be less
active. The mean log ratio is
-2.0, which suggests that new community is usually 13.5\% as
active as the older one. However, a nontrivial fraction of newer communities (25.7\%) are more active.

\para{A closer look at the more active newer communities.}
It's somewhat surprising that 25.7\% of newer communities overtake their
established counterparts. Why does this occur?
\figref{fig:activity_level} presents examples of possible reasons that
the younger community might surpass its older counterpart.

The first reason is that the \modification
represents something that
naturally appeals to more people.  One example
is \communitypair{writers}{fantasywriters}.  As soon as
\communityname{fantasywriters} was created, its activity level was
more than 7 times as great as that in \communityname{writers}.  Here
are top 3 \modifications that consistently lead to more activity: \modifier{the}
(e.g., \communitypair{stopgirl}{thestopgirl}), \modifier{ex} (e.g.,
\communitypair{mormon}{exmormon}), \modifier{steam} (e.g.,
\communitypair{deals}{steamdeals}).

Second, the newer community may be ``equivalent'' to the older one and
the newer one may win the competition.  For example, in the case of
\communitypair{auto}{Autos}, it took \communityname{Autos} a while to
exceed the activity level in \communityname{auto}, but
\communityname{Autos} is now much more popular (see
\figref{fig:activity_case}).

Third, the newer community may actually be the non-modified
one
(14\% of
pairs have this property, as we earlier observed) and the newer one
might achieve popularity because it is more general. For instance,
\communityname{politics} is more popular than
\communityname{ukpolitics} despite the later's earlier founding. In
this case, as soon as \communityname{politics} was created, its
activity level exceeded \communityname{ukpolitics}.

The fourth and relatively rare reason is that the older one may have a
large competitor, in other words, the newer one may originate from an
even bigger community than the older one.  An example is
\communitypair{hiphop}{makinghiphop}.  \communityname{makinghiphop}
started at a similar size as \communityname{hiphop} but exceeded
\communityname{hiphop} significantly later. Although
\communityname{hiphop} and \communityname{makinghiphop} are both
active, there is a much larger hiphop-related community on Reddit,
\communityname{hiphophead}.  \communityname{makinghiphop} might
actually originate from \communityname{hiphophead} instead of
\communityname{hiphop}.

\begin{figure}[t]
    \begin{subfigure}[t]{0.23\textwidth}
        \addExploreFigure{\textwidth}{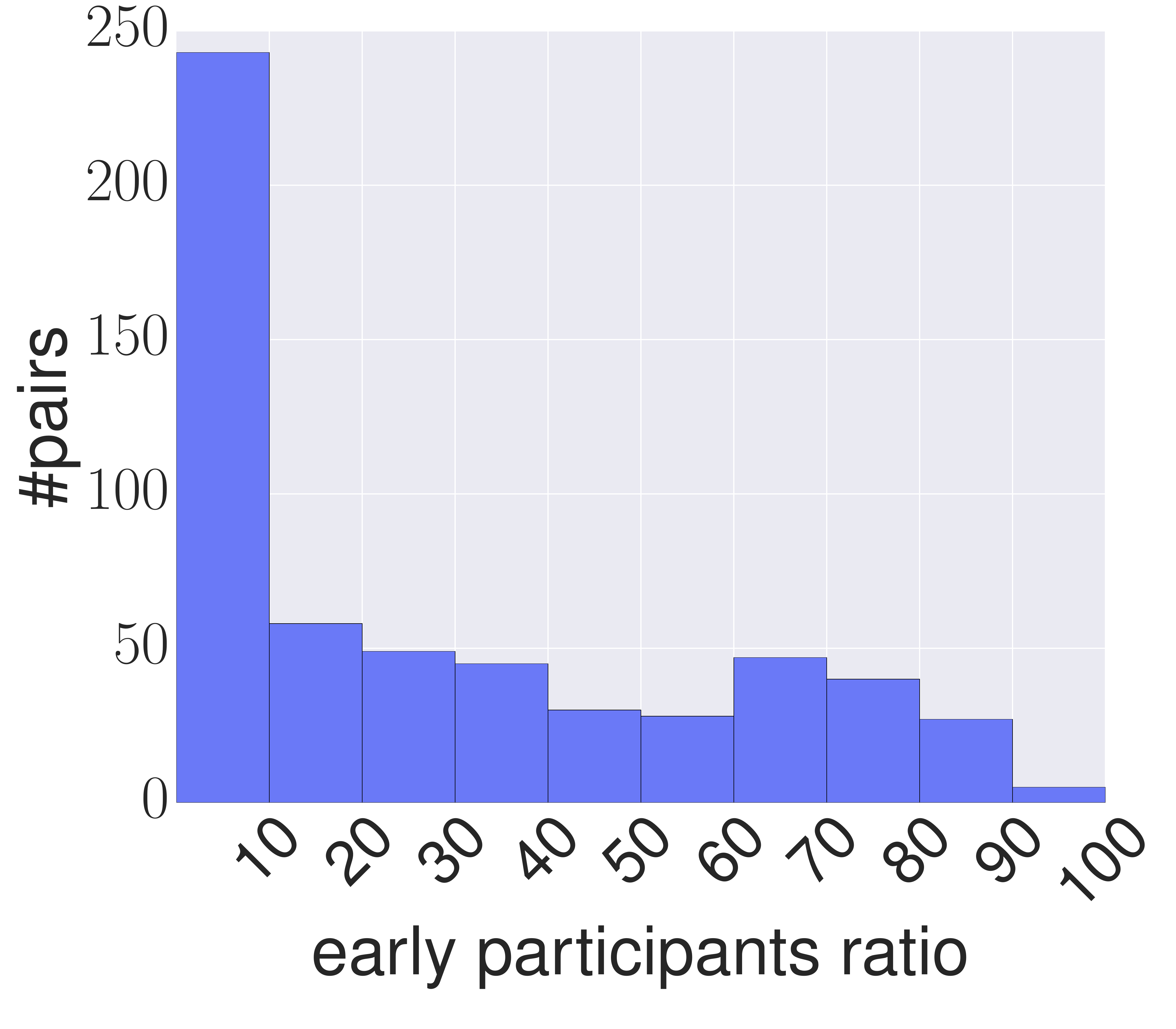}
        \caption{Histogram of the fraction of {\founder}s in the new community that were from the old community over all pairs.}
        \label{fig:founder_dist}
    \end{subfigure}
    \hfill
    \begin{subfigure}[t]{0.23\textwidth}
        \addExploreFigure{\textwidth}{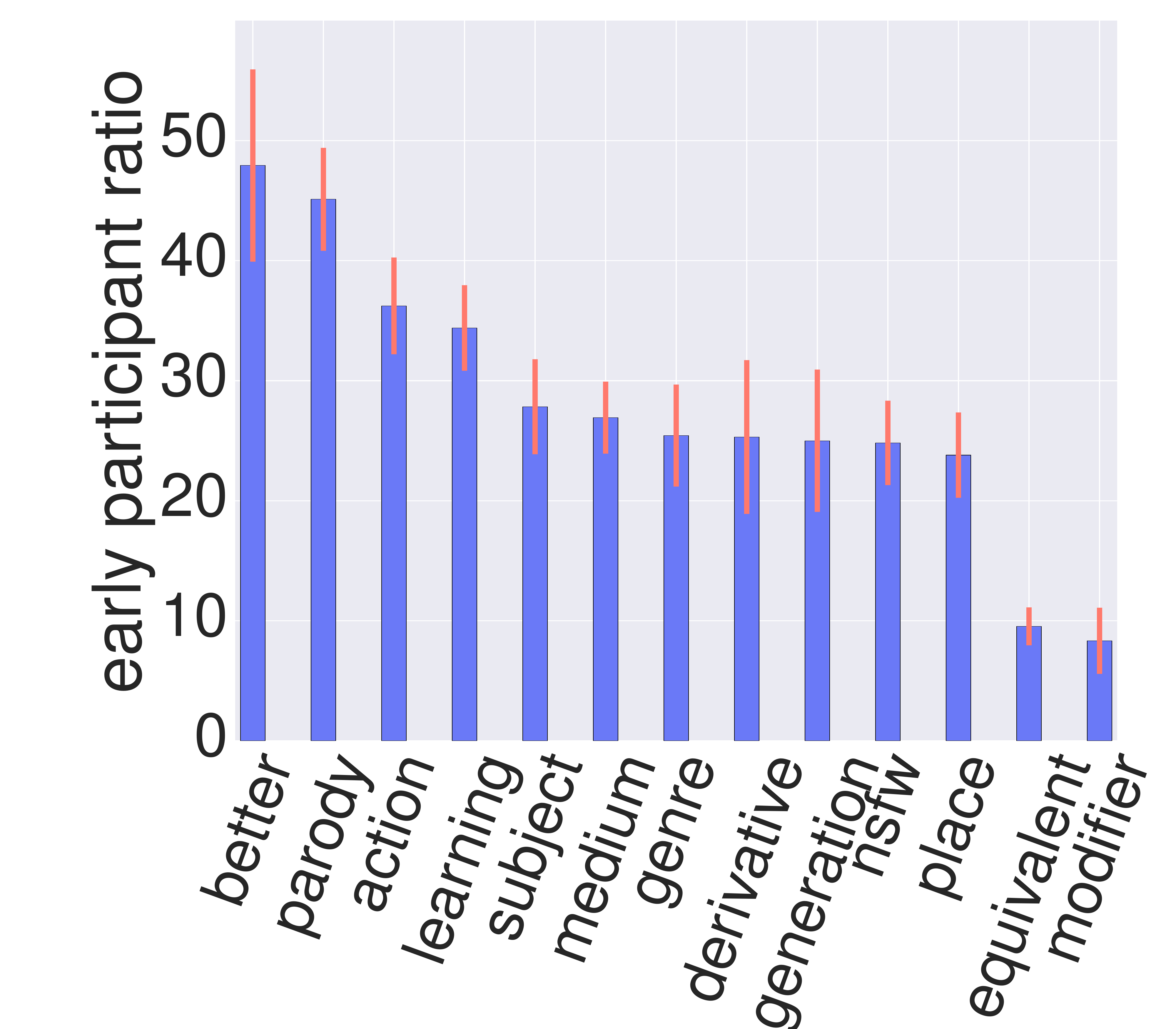}
        \caption{Average fraction of {\founder}s in the new community that were from the old community sorted by categories.}
        \label{fig:top_founder}
    \end{subfigure}
    \caption{
   (a) Surprisingly, the majority of \hrc do not share more than 10\% of {\founder}s.
   (b) \categoryname{Better} has the highest average \founder ratio, while \categoryname{modifier} has the lowest.
    \label{fig:founder}}
\end{figure}

\subsection{Where are early participants in the new communities from?}
The last reason in the above discussion leads to a natural question:
where are the participants in the newer community from?  Are they from
the older one in a pair? This question is difficult to answer, as a
subreddit may establish its own identity and unique audience over time,
even if it was born out of an existing community. If we simply look at
the overlap between two communities over all users, we may mistakenly
believe that they have never shared
the user base as a result of a large number of later users.
We thus focus on
the first $n$ participants in the newer community
(the \emph{{\founder}s}) and compute the fraction of them that were also
members of the older community.
A user is considered a member of the old community if they
took any
action in the old community within the last 30 days prior to
interacting with the new community.
We refer to this metric as
``\founder fraction''.
While we present results for $n=100$, similar results hold for
different $n$.

\para{Almost half of \hrc do not really share {\founder}s.}  As shown
in \figref{fig:founder_dist}, surprisingly, the majority of newer
subreddits in \hrc pairs are not
``founded'' by members of the older
community.
For example,
only 7 of the first 100 participants in
\communityname{makinghiphop}
were members of
\communityname{hiphop}.

\figref{fig:top_founder} presents the
average
\founder ratio
for all categories in \tableref{tb:taxonomy}.  It shows
that \categoryname{better}, \categoryname{parody},
\categoryname{action} and \categoryname{learning} usually attract
members from the older community.  It also partly demonstrates why we
obtain such a low average \founder ration.  \categoryname{equivalent}
and \categoryname{modifier}
appear
more than likely to attract completely
different participants, e.g.,
\communitypair{vegetarian}{vegetarianism}.
We also notice significant differences even within
a single category.  One
notable example is \modifier{meta} (65.8\% from the original
community) vs. \modifier{ex} (1\% from the original community).

\subsection{From \HRC to {\Spinoff}s}
Thus far, we have explored the complex space of possible
\modifications, and the \hrc that are created through them.
We
find
that a non-trivial fraction of the new communities were not
the \modified ones, or did not share the same user base of the
older
one.  For these pairs, it is unclear
whether the new community is a
subdivision of the old one, or whether users in the existing community
are affected by the new one's presence. In order to better understand how
users in the existing community may behave \emph{after exploring the
  new community,} we will focus on a subset of \hrc called
\emph{{\spinoff}s} in the remainder of this work.

\section{{\Spinoff}s: Substitutions or Complements?}
We now formally define \emph{\spinoff} communities.  First: the newer
of the two pairs in a \hrcsingular is a \emph{\spinoff} if it
satisfies the following properties: 1) more than 10\% of the first 100
{\founder}s in the newer community are members of the older community;
and 2) the newer community is the \modified one,
so that it is likely to represent a specialization or some other topic
of interest.
We will sometimes refer to a pair of \hrc that contain a spinoff as a \emph
{\spinoff pair}.

In this section, we investigate
how a user's behavior within the older subreddit is affected once they
try out the newer \spinoff:
do such users get ``distracted'' by the new
one, or does the new
subcommunity complement the old one? Phrased differently, do users tend to
decrease, increase, or not change their activity levels in the
original community after trying the
\spinoff?

Surprisingly, we find that users who \defect the \spinoff generally
become
\emph{more} active in the \emph{original} community.
Furthermore,
with respect to the taxonomy we developed in \tableref{tb:taxonomy},
the magnitude of this trend \emph{depends on the type of
  \modification}: larger in \categoryname{action}, \categoryname{better},
  and \categoryname{parody}, smaller in \categoryname{medium}, and {\em negative} in \categoryname{nsfw}.
Finally, it seems that this complementary effect is more prominent for
users with lower activity level, although there is less data to
compare users with different activity
levels, and results may vary
depending on specific pairs.

\para{Disclaimer:} {\em we do not make any claims of causality given the
observational nature of our dataset.}

\begin{figure}[t]
\centering
\includegraphics[width=.45\textwidth]{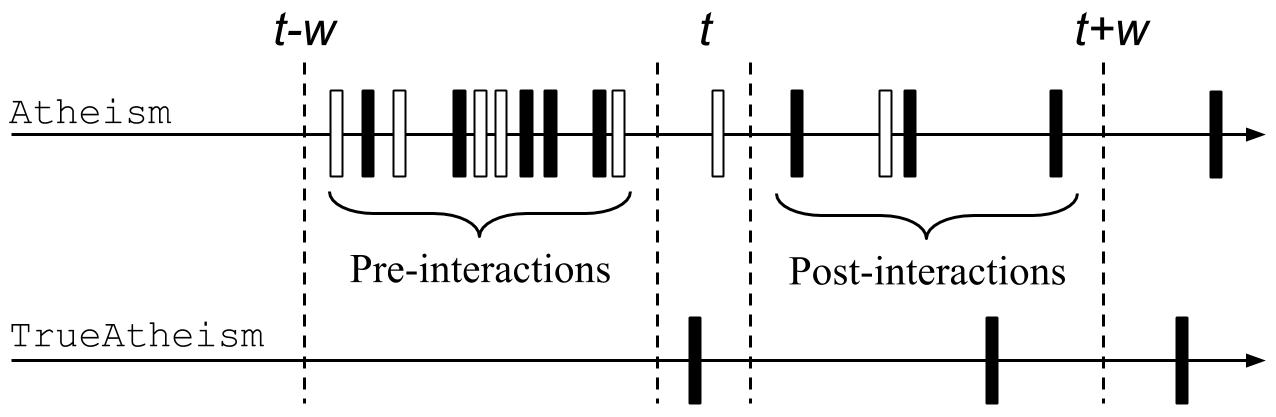}
\caption{
Schematic of the \defection experiment setup. \communityname{TrueAtheism} is a
  \spinoff of \communityname{Atheism}, and the activity of two users is shown over
  time. Each box represents an interaction.
  With respect to the two subreddits shown, the dark user is \adefector,
  and the light user is \aloyalist.
  Time $t$ is the time of the dark user's first
  interaction with the \spinoff subreddit. Here, the number of
  pre-interactions for both the dark and light users is 5. The dark
  user has 3 post-interactions, whereas the light user only has one.}
\label{fig:prepost}
\end{figure}

\begin{figure*}[t]
    \begin{subfigure}[t]{0.45\textwidth}
      \centering
        \addSpinoffFigure{\textwidth}{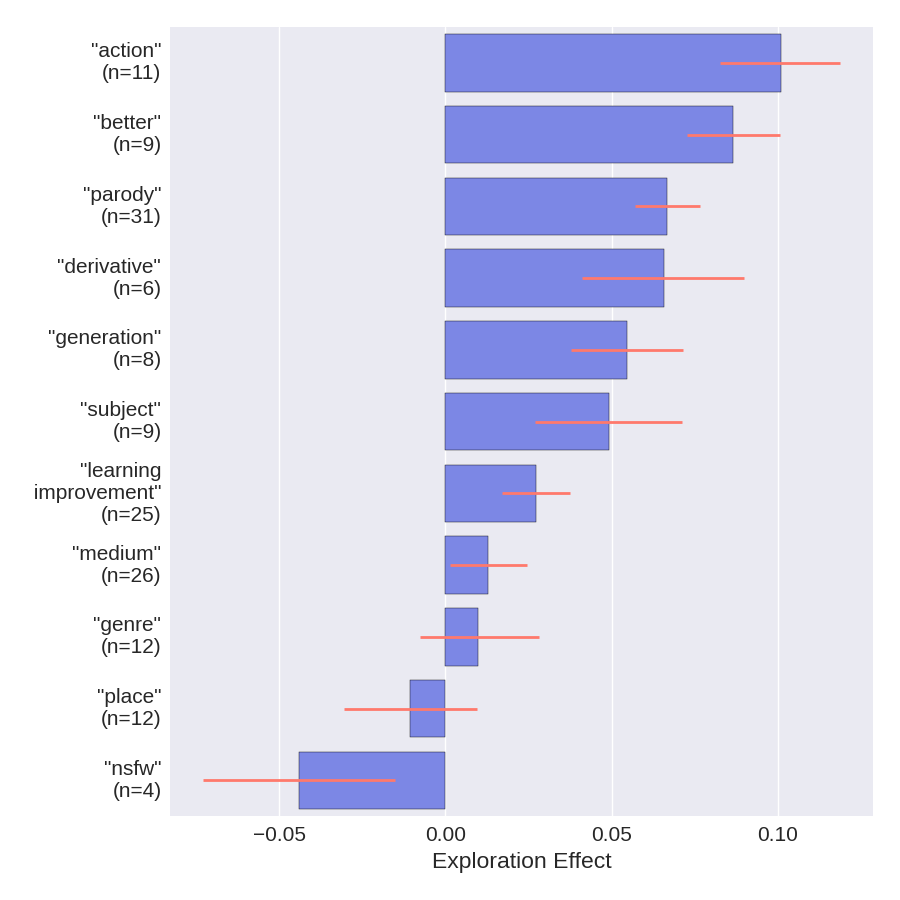}
        \caption{\Metricname by category ($n$ is the number of sampled pairs, only $n \geq 4$ is shown)}
        \label{fig:tax}
    \end{subfigure}
    \hfill
    \begin{subfigure}[t]{0.45\textwidth}
      \centering
        \addSpinoffFigure{\textwidth}{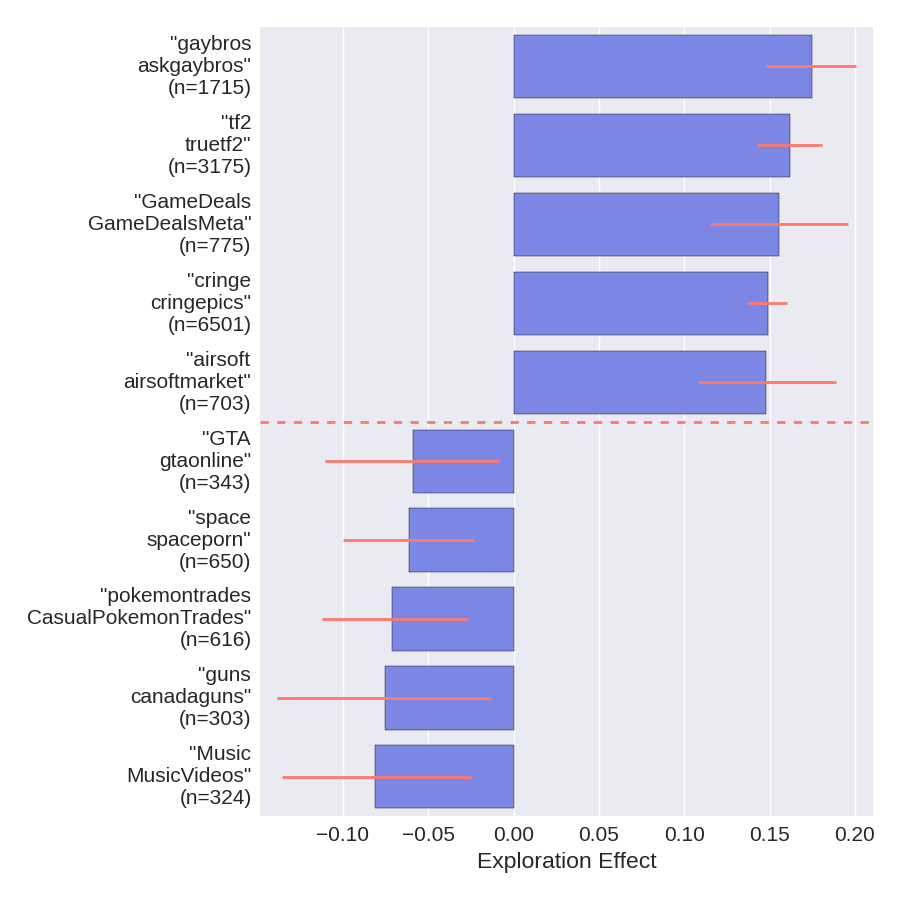}
        \caption{Top/bottom 5 \metricname by pair ($n$ is the number of sampled users)}
        \label{fig:pair}
    \end{subfigure}
    \caption{
    Difference between {\defector}s and {\loyalist}s in the fraction of users that become more active in post-interactions
    (in the older community)
    compared to pre-interactions.
    Larger values indicate more activity from {\defector}s.
      (a)
      categories from our taxonomy
      and (b)
      specific pairs.
      Error bars represent 95\% CIs. }
      \label{fig:splits}
\end{figure*}

\subsection{Experiment setup}
To understand user behavior in the \emph{original} community \emph{after}
participating in the \spinoff community,
we propose an experiment framework
in which we first
pair ``{\adefector}''
and a ``similar'' ``{\loyalist}''
in the original community. After identifying this pair of users, we
compare their behavior pattern after the {\defector} first
participated in the \spinoff community, as illustrated in
\figref{fig:prepost}.

Specifically, for each \spinoff pair (e.g.,
\communitypair{Atheism}{TrueAtheism} in \figref{fig:prepost}),
we define {\em {\defector}s} as users who were active in the original
community in a time window before their first participation in the \spinoff
community.\footnote{Participation and being active are both defined as
  either posting or commenting.}  The darker user in
\figref{fig:prepost} is an example.  We denote the time of her first
interactions in the \spinoff community as $t$, and refer to her
interaction in the original community from $t-w$ to $t$ as {\em
  pre-interactions} and her interaction in the original community from
$t$ to $t+w$ as {\em post-interactions}.  We consider users with at
least 5 pre-interactions to ensure that they were indeed active in the
original community.

A straightforward metric to compute is simply the ratio between the
number of post-interactions and the number of pre-interactions for
each user. However, this is problematic because we require users to
have at least 5 pre-interactions but have no constraints on
post-interactions.  This causes our sample to be biased towards users
with more pre-interactions than post-interactions.

To address this concern, for each \defecting user
$u_{\defsubscript}$,
we sample a \emph{similarly active} user $u_{\loyalsubscript}$ in the
original community who \emph{never} interacts with the \spinoff
community. We call this user a ``{\em \loyalist}''.  The rough idea is
demonstrated by the light user in \figref{fig:prepost}, who had a
similar number of pre-interactions and made a post in the original
community around $t$ so that we know she was still active.  The
details of this sampling process are given in the appendix.

\para{Metric: \metricname.}  After we identify {\defector}s and
matching {\loyalist}s, we compute the fraction of {\defector}s who
have more post-interactions than pre-interactions
in the older community, and then compute the same fraction for {\loyalist}s.
We take the difference between these two fractions and
call it the
``\metricname'' (see Equation \ref{eq:diffdefect}, in the appendix).
Higher values of this quantity
indicate that $u_{\defsubscript}$ was more active in the original
community than the \loyal
$u_{\loyalsubscript}$.
We use the macro average to aggregate results from different \spinoff pairs
because the number of {\defector}s varies between pairs.

The only parameter in our framework is $w$.
Since our primary objective of interest in this work is the effect of the interaction with the \spinoff community,
we choose a relatively small window (30 days) to mitigate confounding factors that may affect user behavior over time and the dynamic nature of online communities \cite{danescu2013no,backstrom2006group,ducheneaut2007life,kairam2012life,kumar2010structure}.
Our results are robust to reasonable changes in $w$ (e.g., $w=20$ days
produces very similar results).

\subsection{More active after \defecting the \spinoff community}

We now apply this framework and examine how {\defector}s behave in
general.  Surprisingly, we find that {\defector}s are relatively more
active compared to {\loyalist}s,
i.e.,
the \metricname is generally
positive.  We then further split {\defector}s based on their activity
level and study how our observation differ depending on activity
level.

\begin{figure*}[ht]
    \begin{subfigure}[t]{0.24\textwidth}
        \addTwoLineFigure{\textwidth}{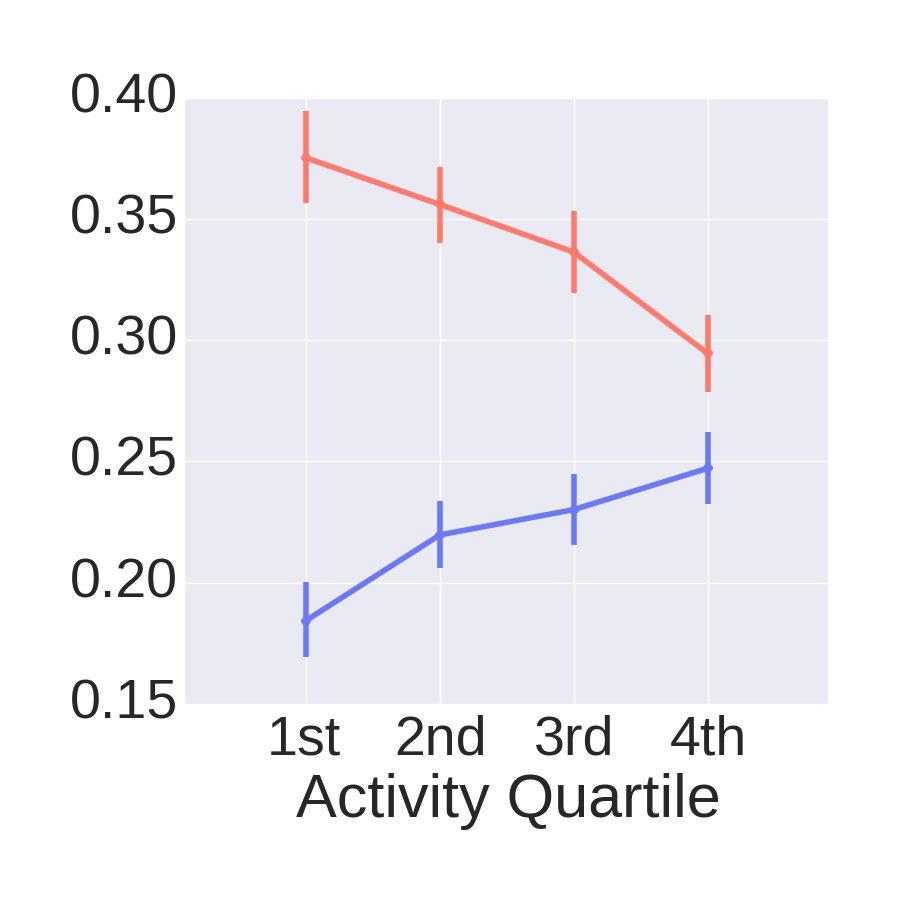}
        \caption{\communityname{AskReddit} vs \communityname{TrueAskReddit}; (n=8816 user pairs)}
        \label{fig:true}
    \end{subfigure}
    \hfill
    \begin{subfigure}[t]{0.24\textwidth}
        \addTwoLineFigure{\textwidth}{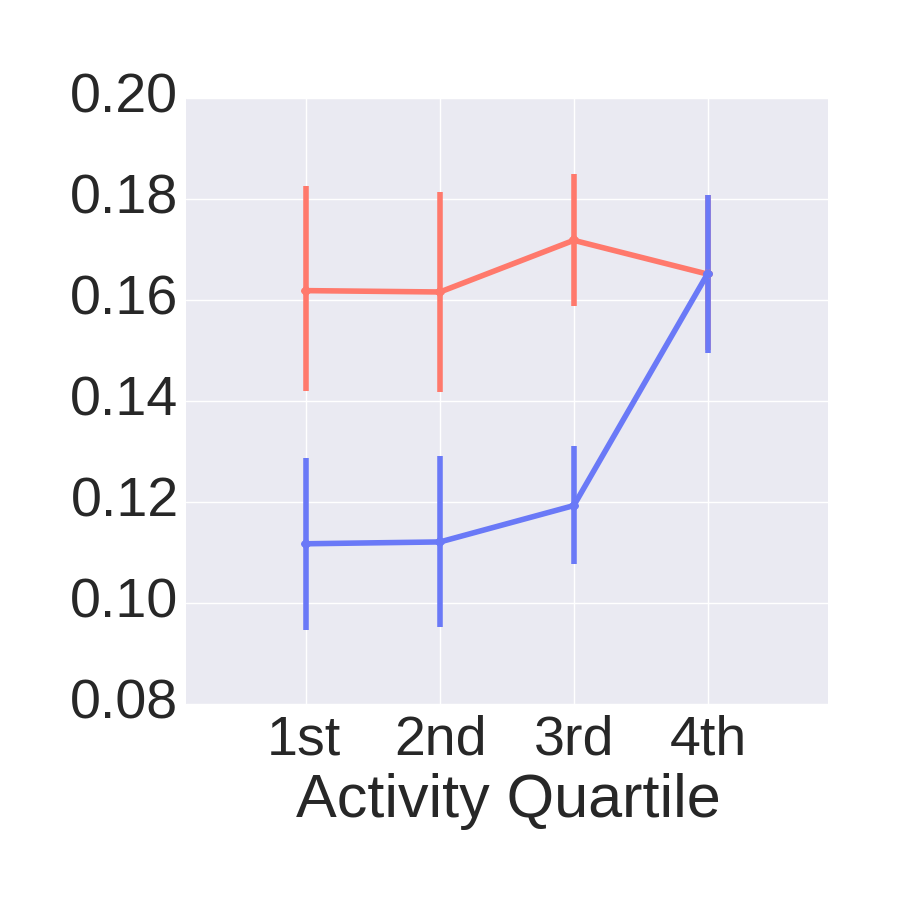}
        \caption{\communityname{Science} vs \communityname{AskScience}; (n=5516 user pairs)}
        \label{fig:ask}
    \end{subfigure}
    \hfill
    \begin{subfigure}[t]{0.24\textwidth}
        \addTwoLineFigure{\textwidth}{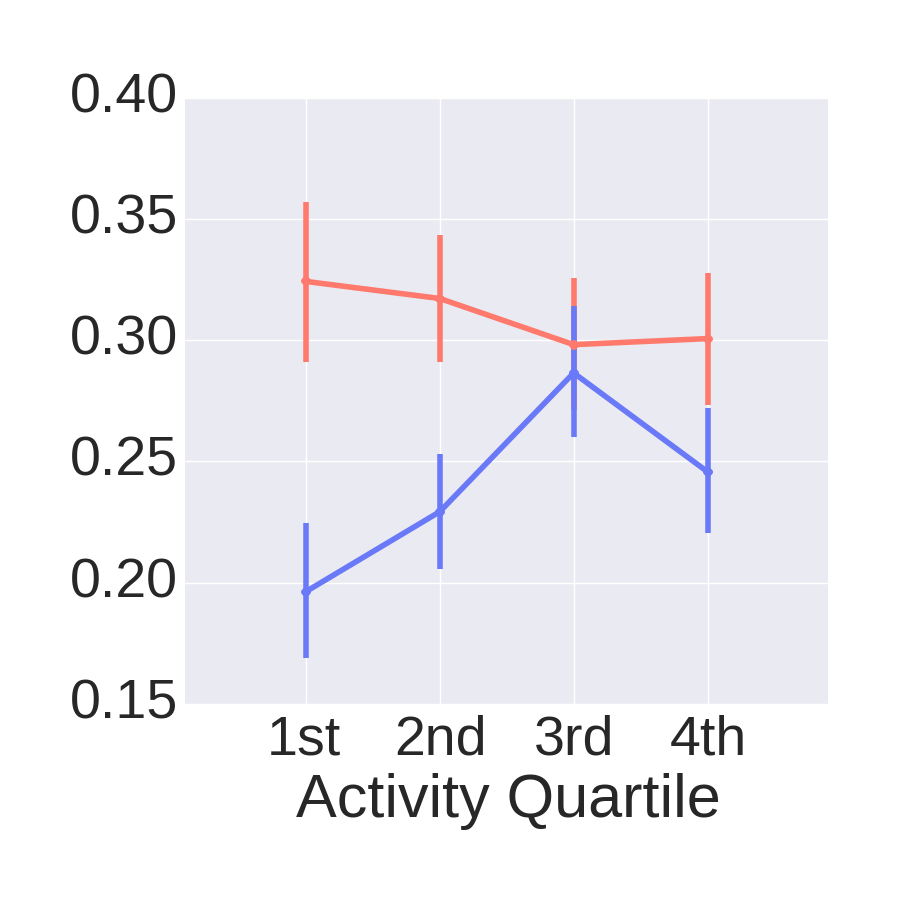}
        \caption{\communityname{Android} vs \communityname{androidapps}; (n=2951 user pairs)}
        \label{fig:apps}
    \end{subfigure}
    \hfill
    \begin{subfigure}[t]{0.24\textwidth}
        \addTwoLineFigure{\textwidth}{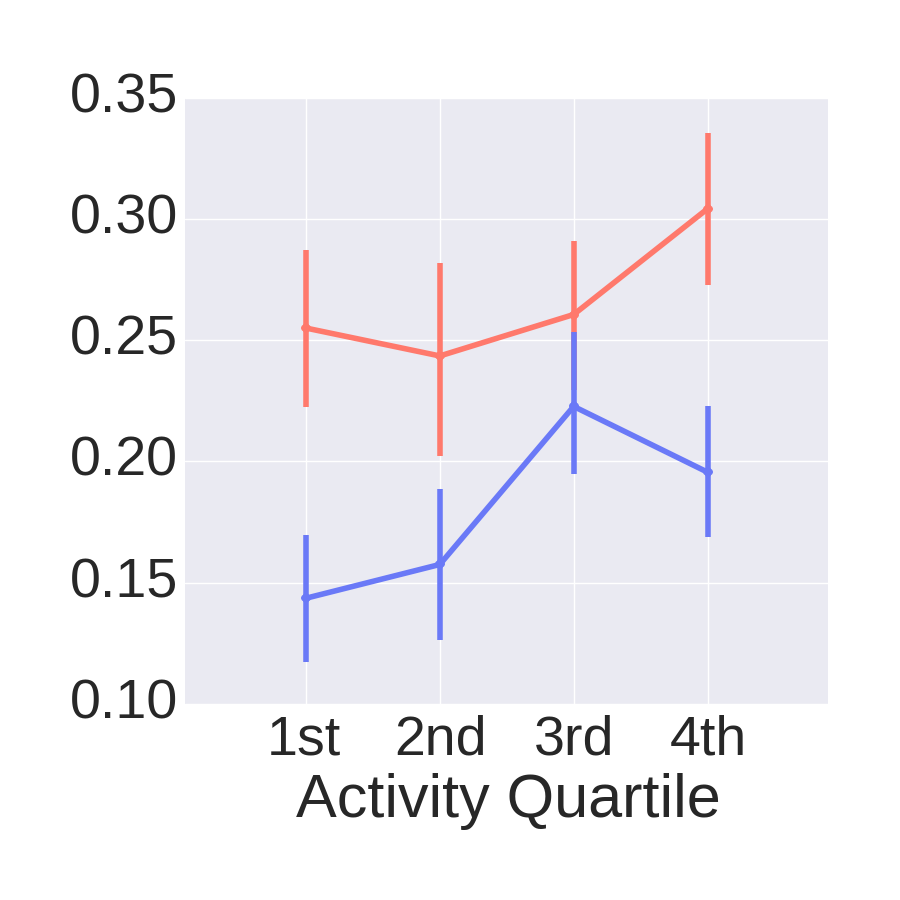}
        \caption{\communityname{apple} vs \communityname{applehelp}; (n=2221 user pairs)}
        \label{fig:help}
    \end{subfigure}
    \caption{Several examples of \textbf{\textcolor{defectorcolor}{\defector}}
    and \textbf{\textcolor{loyalcolor}{\loyalist}} activity levels (with 95\%
    CIs) split into quartiles by pre-activity. The x-axis is
    pre-interaction quartile, and the y-axis is the proportion of
    users whose post-interactions exceeded their pre-interactions. In
    all cases, {\defector}s tend to have greater post-interaction
    levels than {\loyalist}s, reflective of the results from the previous
    section. These plots are meant to highlight the complex
    relationships between activity level and activity rates. We
    observe many statistically significant differences, but note that
    each \spinoff community pair's behavior in this regard appears to be
    unique.
    In the first three pairs, we do see that {\defector}s with the highest pre-activity level present a smaller difference from {\loyalist}s.
    } \label{fig:two-lines}

\end{figure*}

\para{Comparisons across categories.}
\figref{fig:tax} presents \metricname results
for categories in our \tableref{tb:taxonomy} taxonomy.\footnote{Results are
only
reported for categories with more
than 4 \spinoff pairs.
}
Somewhat counterintuitively, we find that for most {\spinoff}s, users
who \defect become \emph{more} active in the original subreddit after
\defecting, compared to similarly active users who never interacted in
the new community (see \figref{fig:tax}).

Interestingly, the magnitude of this result varies based on the
\spinoff pair considered. We observe that \categoryname{action}
{\defector}s are around 10\% more likely to increase their
activity after \defecting, for example.
\categoryname{place}
{\defector}s, on the other hand, are roughly 2\% less likely
to increase their activity.

Our
possible explanation for this observation is that users who \defect\xspace
\categoryname{action} communities are often seeking to actively engage
with a topic in a fashion above and beyond simple discussion. For
example, the subreddit \communityname{Bitcoin} (which focuses on high-level discussions of the crypto-currency) and its \spinoff pair
\communityname{BitcoinMining} (which focuses on lower-level issues,
e.g., hardware useful for mining Bitcoins) exhibits a difference in
interaction ratio of roughly 10\%. If a user {\defect}s
\communityname{BitcoinMining} from \communityname{Bitcoin}, this is
likely a strong indication of their interest in digging deeper into
the topic itself. It's possible that viewing \communityname{Bitcoin}
through the perspective of \communityname{BitcoinMining} increases
overall engagement with the topic, at least in the short term.

In contrast, \defecting
\categoryname{place} subreddits does not
result in increased home activity nearly as often. For example,
\communityname{Bitcoin} has another \spinoff pair,
\communityname{BitcoinUK}, that has \ametricname
of roughly zero. We have previously seen in \figref{fig:top_founder}
that \categoryname{place} {\spinoff}s share relatively few
{\founder}s with their parent communities. Taken together, these
observations suggest
that users seeking place-specific
communities are not necessarily interested in engaging more deeply
with the topic, so much as
in  \emph{who} they discuss the topic with or
\emph{how} the topic affects them.

\para{A closer look at the pairs.} \figref{fig:pair} presents the top and bottom 5 pairs in terms of \metricname.
It further demonstrates how our results may vary across different pairs.
All
5 bottom pairs present significantly negative \metricname, which shows that it is not always the case that {\defector}s are more active.

Looking at the bottom 5, it partly supported our above discussion
regarding places.  Indeed, in \categoryname{place} related pairs,
\communityname{gtaonline}
pulled
people from \communityname{gta}
and so did \communityname{canadaguns} for \communityname{guns}.  Among
the top 5, there is an even spread among several categories including \categoryname{learning}
(\communitypair{gaybros}{askgaybros}), \categoryname{action} (\communitypair{airsoft}{airsoftmarket}),
and \categoryname{medium} (\communitypair{cringe}{cringepics}).  The surprising
\modification is \modifier{true}.  Although it seems to suggest
superiority and separation, {\defector}s actually become more active
in the original community in this case, too.

\para{Discussion.}
Our findings resonate with \citeauthor{tan2015all}'s
\shortcite{tan2015all} results that users who continually
explore new communities are, on average, more active than users who
don't.
{However,
no causal relationship can be established that explains this result%
:
\defection does not necessarily {\em cause} increased activity.
However, in our dataset, \defection appears to be a strong signal of
interest level.}

\subsection{Variations between {\defector}s with different activity levels}

We have established that users tend to
be relatively more active
in the ``older'' community
after \defection, and have
examined the
variation across different categories and pairs.  However, how does
this effect differ for users with different pre-interaction levels?
One could imagine that activity level
prior to \defection affects whether or not users are more active after
\defecting.  For example, upon discovering an alternative community,
it's possible very active users might remain more
attached
to their home community, whereas relatively inactive users
might not have the same level of commitment.

To address this question, we
split users into pre-interaction quartile levels within their \spinoff
pair, so that the users with the least number of pre-interactions are
put in bin one,
and users with the greatest number of pre-interactions are put in bin
four.
We then compute \metricname for users in each
quartile.\footnote{We have previously referred to Equation
  \ref{eq:diffdefect} as \metricname, but plot $p_{\defsubscript}$ and
  $p_{\loyalsubscript}$ separately in these plots under the same
  name.}
Figure \ref{fig:two-lines} presents the fraction of users who had more
post-interactions than pre-interactions for, respectively,  {\defector}s
and {\loyalist}s in several popular subreddit pairs.  In general, the
relative effects of \defection appear to be different based on how
active users are, but there are complex and varied relationships
between user activity level and how much defection matters; these
relationships differ based on which \spinoff pair is considered.
Since we split users further into quartiles, the amount of data is not
sufficient to reach conclusions for all pairs.

One relatively {\em consistent} pattern across pairs is that
{\defector}s with the highest pre-activity level usually have a
smaller difference from the {\loyalist}s compared to {\defector}s in
the lowest quartile, as shown in the left three figures in
\figref{fig:two-lines}, although this is not true for
\figref{fig:help}.

The trend of how the fraction or the difference changes with different
pre-activity quartile is even more complex.  Consider the case of
\figref{fig:true}; this figure illustrates that for users with low
activity levels (first/second quartiles) \defecting is much more
indicative of increased future activity than not \defecting,
and the difference is much less apparent for users with high activity levels -- \defecting and not \defecting are associated with more
similar levels of activity for users in the third/fourth quartiles.
Note that other pairs exhibit different patterns. For \communityname{Science}
vs \communityname{AskScience} (\figref{fig:ask})
and \communityname{Android} vs \communityname{androidapps}
(\figref{fig:apps}), the most active users (those in the $4^{th}$
quartile) appear to experience a slight ``dip'' in terms of the \metricname.

\section{Related Work}

While there has been considerable interest in the topic
in
the social sciences
(e.g., \namecite{hurtado1997understanding,berry1997immigration}), the study
of situations wherein users engage with \emph{multiple}, distinct
communities represents a relatively new but increasingly relevant
research area for computer scientists. Indeed, Kim
\shortcite{Kim:2000:CBW:518514} argues that a growing Web \emph{needs}
subdivisions, while Jones and Rafaeli
\shortcite{Jones:ElectronicMarkets:2010} also argue that an effective
community splitting strategy is essential for virtual communities and
online discourse to thrive. Furthermore, Birnholtz et al.'s
\shortcite{Birnholtz:2015:WSV:2702123.2702410} study of confession
groups on Facebook could be viewed in the context of ``place'' style
{\modification}es. While our study has a different focus, our findings
mirror those of Tan and Lee \shortcite{tan2015all} who found that
users who tend to remain active on the Reddit platform (as an example of a meta-community)
tend to continue to explore new sub-communities continuous throughout their
``lifetime'' on the site.

A number of studies have examined multi-community platforms in
different contexts. Subcommunity survival
\cite{turner2005picturing,iriberri2009life,kraut2012building} is
sometimes framed in the context of a meta-community. Also, Fisher et
al. \shortcite{fisher2006you} find that different newsgroups exhibit
different conversation patterns, though
tehy
don't examine if the same
users behave differently across platforms (as in
\namecite{vasilescu2013stackoverflow}). Finally,  \namecite{adamic2008knowledge} examine the quality of user
answers across different categories of Yahoo Answers.

Despite exhibiting some undesirable upvoting patterns
\cite{gilbert2013widespread}, Reddit itself has been used as a data
source in various contexts. For instance,
the study of
altruistic requests \cite{althoff+al:14a},
the study of domestic abuse discourse \cite{ray4183analysis}, and
work about \submission
titles
\cite{lakkaraju2013s}
demonstrate that useful information can be
learned from Reddit comments and upvotes.

\section{Conclusion}

In this work, we
use
a dataset of all {\submission}s and comments
from Reddit over an eight-year period to explore the space of naming
\modifications that lead to \hrc on the platform.
After building a taxonomy,
we
examine
the
{\founder}s and other temporal aspects of the pairs, and
introduce
the idea of a \spinoff community being ``born
out'' of its \unmodified parent. Finally, we
present
the surprising result that users who \defect in \spinoff
communities generally
become relatively \emph{more} active in their home communities
instead of being
``distracted''.
We also
find that the magnitude of this effect (and sometimes its sign)
depend on the type of
community pair
and how active a user was prior
to \defection.

There are several directions for possible future work. First, it would
be interesting to examine more closely the \emph{origins} of \hrc. If
a community is created because of a disagreement (e.g.,  Zachary's
Karate Club \shortcite{zachary1977information}) one could potentially
identify general characteristics of increasing unrest prior to a
fission. Also, it would be interesting to delve deeper into
differences between discourse on content in \hrc pairs; how does
discussion on \communityname{TrueAtheism} differ from discourse on
\communityname{Atheism}, for example.
It would be useful for community organizers if we can detect when a \spinoff community is necessary or beneficial.
Furthermore, it is an important direction to understand the mechanism behind
our observation that users who \defect in \spinoff
communities generally
become relatively \emph{more} active in their home communities.
This could be potentially useful for community organizers to identify complementary communities.

Finally, we note that our consideration has presupposed a pairwise
framing, i.e., we always assumed a \emph{pair} of communities. In some
cases, we noted more complex phenomena underlying community
creation. For example,
a number of \communityname{srs} communities were all created in a short period of time.
Also, the world of \communityname{pokemon} subreddits may consist of multiple \modifications that lead to different subdivisions.
In general, one could generalize pairwise interactions to
explore more complex
relationships between communities.

\para{Acknowledgments.}
We thank
Cristian Danescu-Niculescu-Mizil,
David Mimno,
Mor Naaman,
Skyler Seto,
Tianze Shi,
Justine Zhang
and the anonymous reviewers
for helpful comments and discussions.
This work was supported in part by a Facebook fellowship
and a Google Research Grant.

\section{Appendix: Sampling Method for Control Users}

The goal of this section is to describe how we
sample a control
user $u_{\loyalsubscript}$ corresponding to each each \defecting user
$u_{\defsubscript}$. Ultimately, to compute the \metricname, we need
to find someone who never posts in the new subreddit, but
takes a similar number of actions in the same time period. To choose this similarly
active, \loyal user, we sample $u_{\loyalsubscript}$ as follows:
\begin{enumerate}
\item From the set of all \loyal users, find the subset who have an
  interaction in the original community within 24 hours of
  $u_{\defsubscript}$'s \defection time $t$. Let these interactions
  occur at time $t'$. If a \loyal user has more than one interaction
  between $t-24$ hours and $t+24$ hours, take the closest to $t$.
\item Find the user $u_{\loyalsubscript}$ in this candidate set that
  minimizes the difference between their own number of
  pre-interactions (re-centered at their $t'$) and
  $u_{\defsubscript}$'s. Specifically, if we let $p(u, t_a, t_b)$ be the number of
  interactions of user $u$ in the original subreddit between $t_a$ and
  $t_b$ we find the loyal user
  $$ \argmin_{u_{\loyalsubscript}} | p(u_{\defsubscript}, t-w, t) - p(u_{\loyalsubscript}, t'-w, t') | \, .$$
\item If this difference is less than 5\% of $u_{\defsubscript}$'s pre-interactions,
  a similarly active user $u_{\loyalsubscript}$ has been successfully sampled.
\end{enumerate}
Figure \ref{fig:prepost} demonstrates a pair of users that could be
plausibly sampled in this manner. Both the light and dark users have
the requisite 5 pre-interactions, and the light user makes a post
within 24 hours of the dark user's first \defection.

After sampling $k$ such user pairs $\{\langle u_{i,\loyalsubscript},
u_{i,\defsubscript} \rangle\}|_{i=1}^k$ for a given pair of subreddits%
\footnote{We discard the pair of communities if $k < 100$.}, we first compute the
proportion of \defecting/\loyal users whose activity increased,
i.e., have more post-interactions than pre-interactions.  For
instance, this fraction for \defecting users is computed as
\begin{equation}
p_{\defsubscript} := \frac{1}{k}\sum_{i=1}^k
\mathbb{1}
\left[post(u_{i,\defsubscript}) > pre(u_{i,\defsubscript})  \right] \, .
\label{eq:aveprop}
\end{equation}
Finally, for each \spinoff pair of communities, the quantity we are
interested in is
\begin{equation}
p_{\defsubscript} - p_{\loyalsubscript} \, .
\label{eq:diffdefect}
\end{equation}
We generally call
the quantity given in Equation \ref{eq:diffdefect} the ``\metricname''.
A larger
\metricname indicate that \adefector is more active in the
\emph{original} subreddit after posting to the splinter subreddit,
when compared to a similarly active \loyalist. In \figref{fig:two-lines},
we plot $p_{\defsubscript}$ and $p_{\loyalsubscript}$ separately, whereas
in \figref{fig:splits} we plot $p_{\defsubscript} - p_{\loyalsubscript}$.

\small
\bibliographystyle{aaai}
\bibliography{refs}

\end{document}